\documentclass[aps,prl,groupedaddress,superscriptaddress,twocolumn,showpacs]{revtex4}
\usepackage{graphicx}
\usepackage{amssymb}

\begin{document}

\title{Tunneling Spectroscopy and Vortex Imaging in Boron-Doped Diamond}

\author{B. Sac\'{e}p\'{e}}
\email[electronic adress : ]{benjamin.sacepe@cea.fr}
\affiliation{DRFMC-SPSMS, CEA Grenoble, 17 rue des Martyrs, 38054 Grenoble Cedex 9, France}
\author{C. Chapelier}
\affiliation{DRFMC-SPSMS, CEA Grenoble, 17 rue des Martyrs, 38054 Grenoble Cedex 9, France}
\author{C. Marcenat}
\affiliation{DRFMC-SPSMS, CEA Grenoble, 17 rue des Martyrs, 38054 Grenoble Cedex 9, France}
\author{J. Ka\v{c}mar\v{c}ik}
\affiliation{DRFMC-SPSMS, CEA Grenoble, 17 rue des Martyrs, 38054 Grenoble Cedex 9, France}
\affiliation{Laboratoire d'Etude des Propri\'{e}t\'{e}s Electroniques des Solides, CNRS, B.P. 166, 38042 Cedex 9, France}
\affiliation{Centre of Low Temperature Physics IEP Slovakian Academy of Sciences \& FS UPJ\v{S}, Watsonova 47, 043 53 Ko\v{s}ice, Slovakia}
\author{T. Klein}
\affiliation{Laboratoire d'Etude des Propri\'{e}t\'{e}s Electroniques des Solides, CNRS, B.P. 166, 38042 Cedex 9, France}
\author{M. Bernard}
\affiliation{Laboratoire d'Etude des Propri\'{e}t\'{e}s Electroniques des Solides, CNRS, B.P. 166, 38042 Cedex 9, France}
\author{E. Bustarret}
\affiliation{Laboratoire d'Etude des Propri\'{e}t\'{e}s Electroniques des Solides, CNRS, B.P. 166, 38042 Cedex 9, France}

\date{\today}

\begin{abstract}
We present the first scanning tunneling spectroscopy study of single-crystalline boron doped diamond. The measurements were performed below 100 mK with a low temperature scanning tunneling microscope. The tunneling density of states displays a clear superconducting gap. The temperature evolution of the order parameter follows the weak coupling BCS law with $\Delta(0)/k_B T_c \simeq 1.74$. Vortex imaging at low magnetic field also reveals localized states inside the vortex core that are unexpected for such a dirty superconductor.
\end{abstract}

\maketitle

	The recent discovery of a superconducting transition in boron-doped diamond provides a new interesting system for the study of superconductivity in doped semiconductors \cite{Ekimov}. In the broad band gap of diamond ($\sim 5.5\,eV$), boron impurities introduce an acceptor level with a hole binding energy of $\sim 0.37\, eV$ and lead to a metallic state above a critical boron concentration in the range of a few atoms per thousand. According to the doping level, the superconducting transition is observed with a critical temperature varying between $1K$ and $10K$ \cite{Bustarret,Umezawa2}. 
 
Because of the strong bonding states and the hole doping, various theoretical studies have stressed the similarity between diamond and the superconductor $MgB_2$ \cite{Boeri,Lee,Blase}. The origin of the remarkable superconductivity in $MgB_2$ at $T_c\,=\,39K$ is now well understood to be a result of phonons coupled to holes in the two dimensional $\sigma$-bonding valence bands. Although the band structure of diamond is three dimensional, the superconductivity in boron doped diamond could also be due to the coupling of a few holes at the top of the valence band to the optical bond-stretching zone-center phonons \cite{Boeri,Lee} as well as to boron-related modes \cite{Blase,Xiang}. Numerical calculations of the electron-phonon coupling by the density-functional supercell method are claimed to yield a calculated $T_c$ in good agreement with experiments for at least one boron concentration \cite{Xiang}, thus supporting that boron-doped diamond is a phonon mediated superconductor. 

	Another theoretical approach proposed by Baskaran \cite{Baskaran1} considers diamond, a broad band insulator, to be an appropriate vacuum state for the boron subsystem; electron correlations then drive a Anderson-Mott insulator to resonating valence bond superconductor transition \cite{Baskaran3}. Because of the proximity to the boron critical density of the insulator-metal transition, the superconductivity takes place within the impurity band. In the disordered lattice of boron acceptors, such a strongly correlated impurity band is expected to give rise to an extended s-wave superconducting state.

	In this letter we report on low temperature tunneling spectroscopy and vortex images of superconducting hole-doped diamond films. These measurements which probe the quasiparticule excitations near the Fermi energy, show a clear BCS Local Density of States (LDOS) consistent with weak coupling. Contrary to what is expected for a dirty superconductor \cite{Golubov,Renner}, a significant density of localized resonant states is found below the gap in the vortex core.

\begin{figure}[h]
 \includegraphics[width=0.3\textwidth]{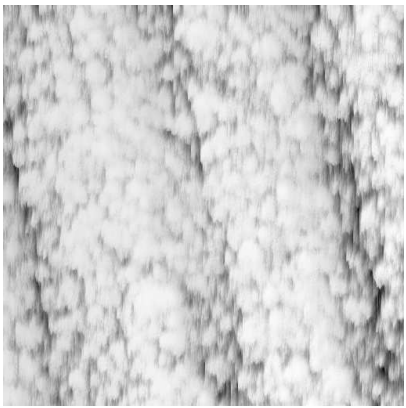}
 \caption{\label{Topo}Topography image ($1.5 \times 1.5 \; \mu m^2$) of the sample obtained at $70\; mK$. The rms rugosity is $1.8\,nm$.}
\end{figure}

	The sample studied was a selected boron-doped single-crystalline diamond epilayer synthesized at $820^oC$ by microwave plasma-assisted decomposition (MPCVD) of a gas mixture composed of Hydrogen, Methane and Diborane with a controlled boron to carbon concentration ratio. A 001-oriented type Ib diamond substrate was exposed to a pure hydrogen plasma prior to diborane-free growth of a $500\,nm$-thick buffer layer. Optical and transport measurements showed that these steps were critical for achieving a high quality boron-doped  diamond epilayer. A boron atomic density varying between $n_B = 3.1\times 10^{21}\;cm^{-3}$ at the surface and $n_B = 2.1\times 10^{21}\;cm^{-3}$ at the interface, as well as an epilayer thickness of $75\; nm$ were derived from a quantitative comparison to a Secondary Ion Mass Spectroscopy (SIMS) profile measured in a B-implanted diamond crystal with a known peak boron concentration. Because the density and effective masses of the charge carriers are not clearly established \cite{Bustarret2}, transport caracteristics such as the Fermi velocity $v_F$ and mean free path $l$, can only be roughly estimated. For a carrier density deduced from the boron density and a sheet resistance of $83\, \Omega/_{\square}$ at $4.2\,K$, $l$ is estimated within the free electron model to be less than $1.3\, nm$. The surface topography obtained with the scanning tunneling microscope (STM) is displayed in Fig. \ref{Topo}. The parallel $600\; nm$ wide  stripes reflect the vicinal surface structure of the substrate and can be observed everywhere with the same orientation. The rms micro-roughness is $1.8\,nm$ and the in plane characteristic size of the surface corrugation is about $75\pm 25\; nm$.

\begin{figure}
 \includegraphics[width=0.45\textwidth,bb = 0 0 530 665]{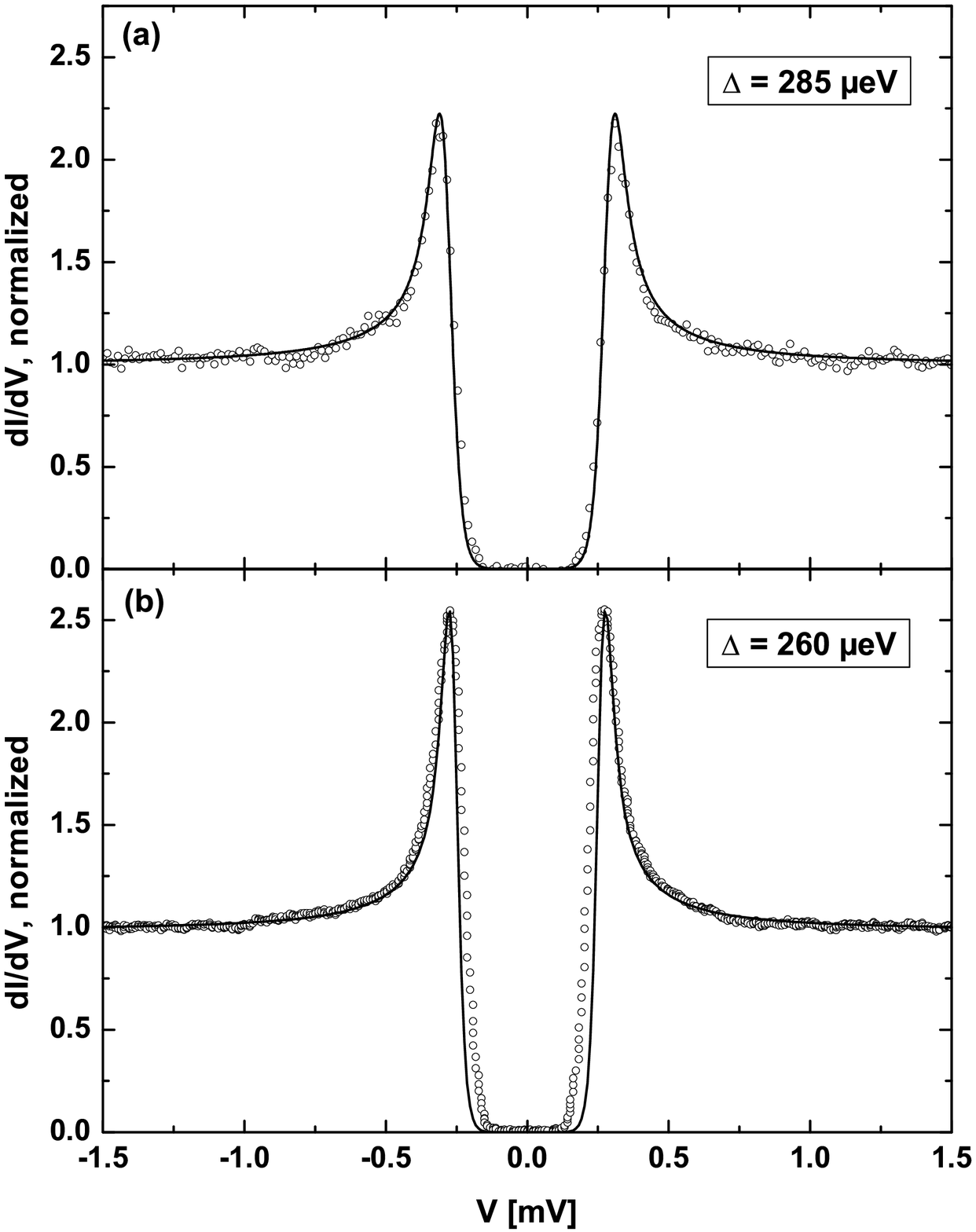}
 \includegraphics[width=0.45\textwidth,bb = 15 15 282 215]{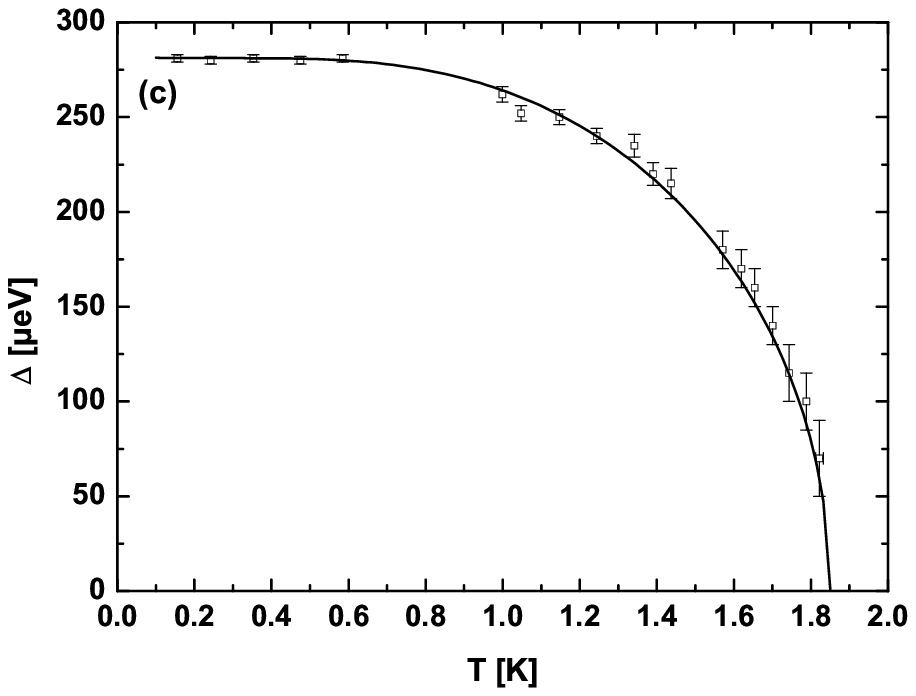}
 \caption{\label{gap} Top : experimental normalized tunneling conductance measured at $70\;mK$ (open circle). Solid lines correspond to BCS fits with : (a) $\Delta = 285\; \mu eV$ and $T_{eff}=235\;mK$; (b) $\Delta = 260\; \mu eV$ and 
$T_{eff}=175\;mK$. Bottom : (c) temperature dependance of the BCS gap (open squares) compared with a BCS law with $T_c = 1.85\; K$ 
(solid line).}
\end{figure}

	Spectroscopic measurements were performed with a home-built Scanning Tunneling Microscope (STM)cooled down to a base temperature of $50\; mK$ in a dilution refrigerator. The sample was scanned with a Pt-Ir cut tip. In order to probe the LDOS, a small AC modulation of $10\;\mu V$ rms was added to the sample-tip DC bias voltage V and the differential conductance $G(V)=\frac{dI}{dV}(V)$ obtained with a lock-in amplifier. At any position above the sample, V could be ramped and the resulting $G(V)$ curve giving the LDOS around the Fermi energy measured. In addition, a static magnetic field perpendicular to the sample surface could be applied with a superconducting coil surrounding the STM. 

	Tunneling spectroscopy performed at different locations revealed the surface to be superconducting with very little spatial inhomogeneity. In Fig. \ref{gap} we show two representative differential conductance curves performed at $70\;mK$ with a tunnel 
resistance of $20\; M\Omega$ and without any averaging. Most of the experimental spectra can be well reproduced by a theoretical BCS density of states as can be seen in Fig. \ref{gap} (a). The fit gives a superconducting order parameter $\Delta = (\,285\pm 2\,)\,\mu eV$. A thermal broadening with an effective temperature $T_{eff} = 235\; mK$ in the LDOS calculation indicates that the energy resolution of our STM is probably limited by unfiltered electromagnetic radiation which heats the electron bath. No additionnal parameters to describe pair breaking were needed. Although the spectrum was almost identical everywhere on the scanned surface, we also obtained smaller values of $\Delta $ in several independent experiments on this sample (for example, see Fig. \ref{gap} (b)). This slight dispersion of the gap value can be either the result of spatial macroscopic inhomogeneity of the doping or be the consequence of variation in the chemical cleaning of the surface prior to each run. The temperature dependence of the order parameter $\Delta (T)$ displayed in Fig. \ref{gap} is well described by the BCS theory with a critical temperature of $1.85\,K$ very close to the superconducting transition temperature of $1.9\,K$ obtained by ac susceptibility and transport measurements. This clearly demonstrates that boron-doped diamond is well described by s-wave BCS superconductivity with a measured ratio $\frac{\Delta}{k_B T_c}\simeq 1.74$ characteristic of weak-coupling \cite{Mitrovic}, as expected from theoretical calculations \cite{Boeri,Blase,Lee,Xiang}. Nevertheless we also found a small number of spectra as shown in Fig \ref{gap} (b) which present broader gap edges. In order to explain them, further investigations of the LDOS for different crystallographic orientations are needed to investigate a possible anisotropy of the order parameter.

\begin{figure}
 \includegraphics[width=0.238\textwidth]{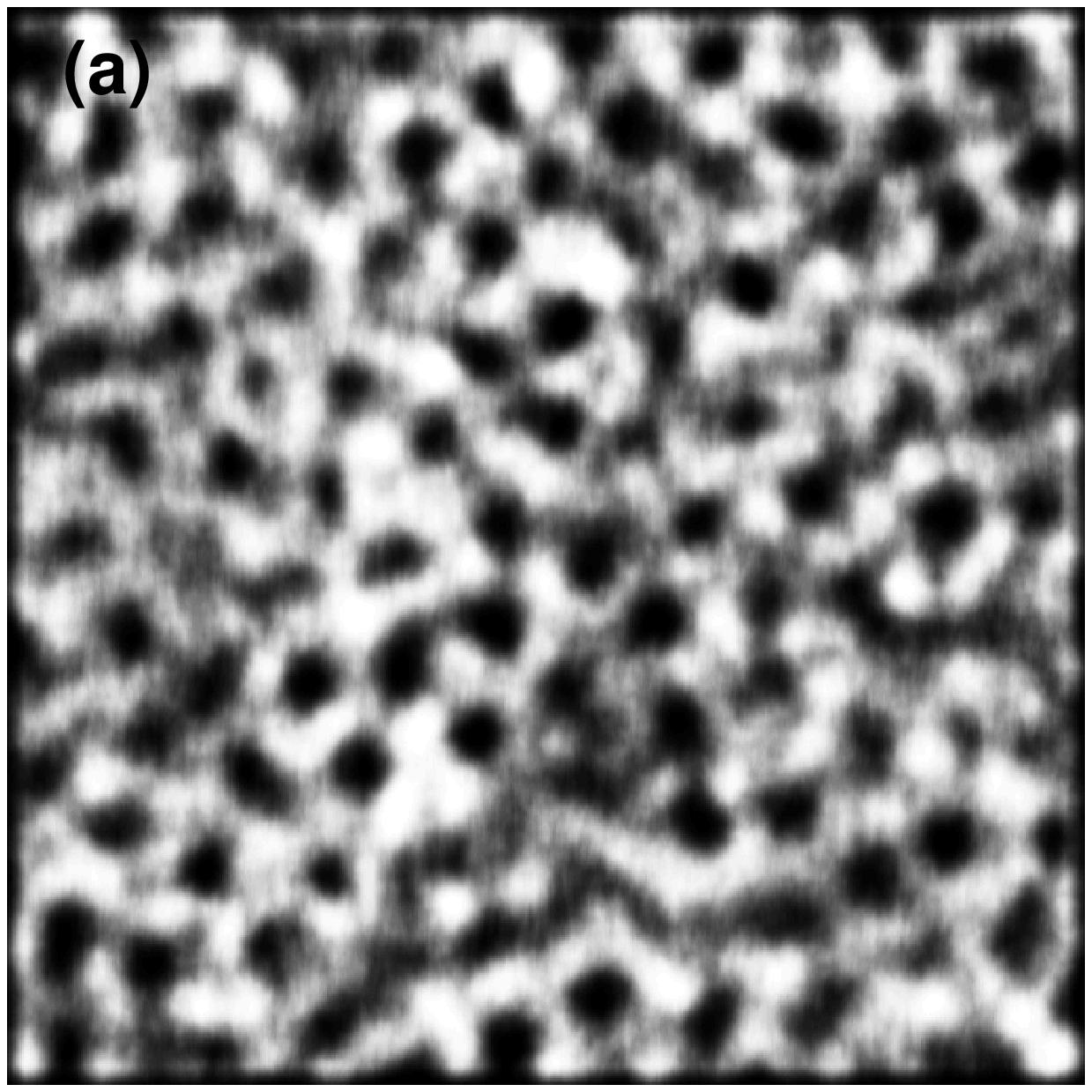}
 \includegraphics[width=0.238\textwidth]{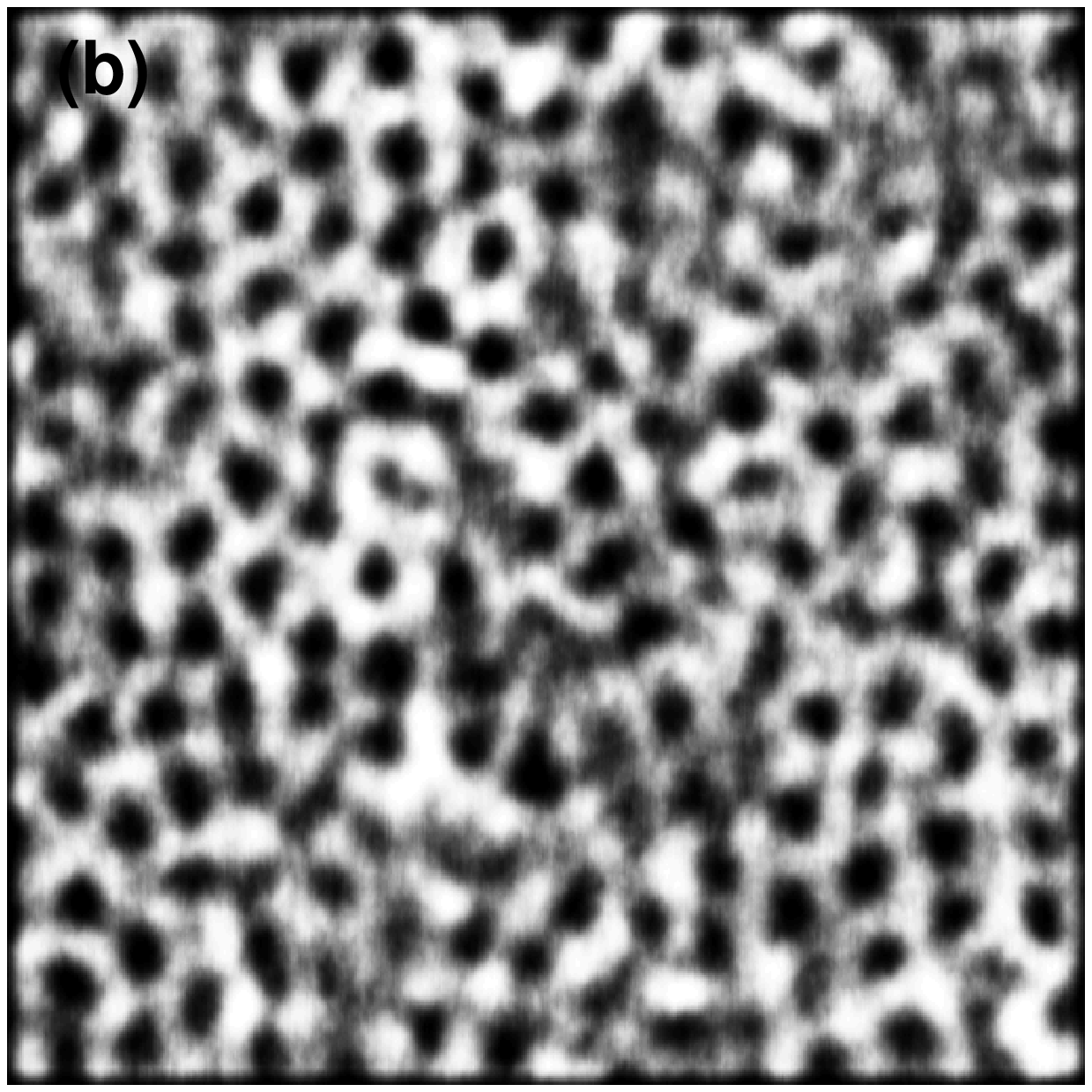}
 \includegraphics[width=0.238\textwidth]{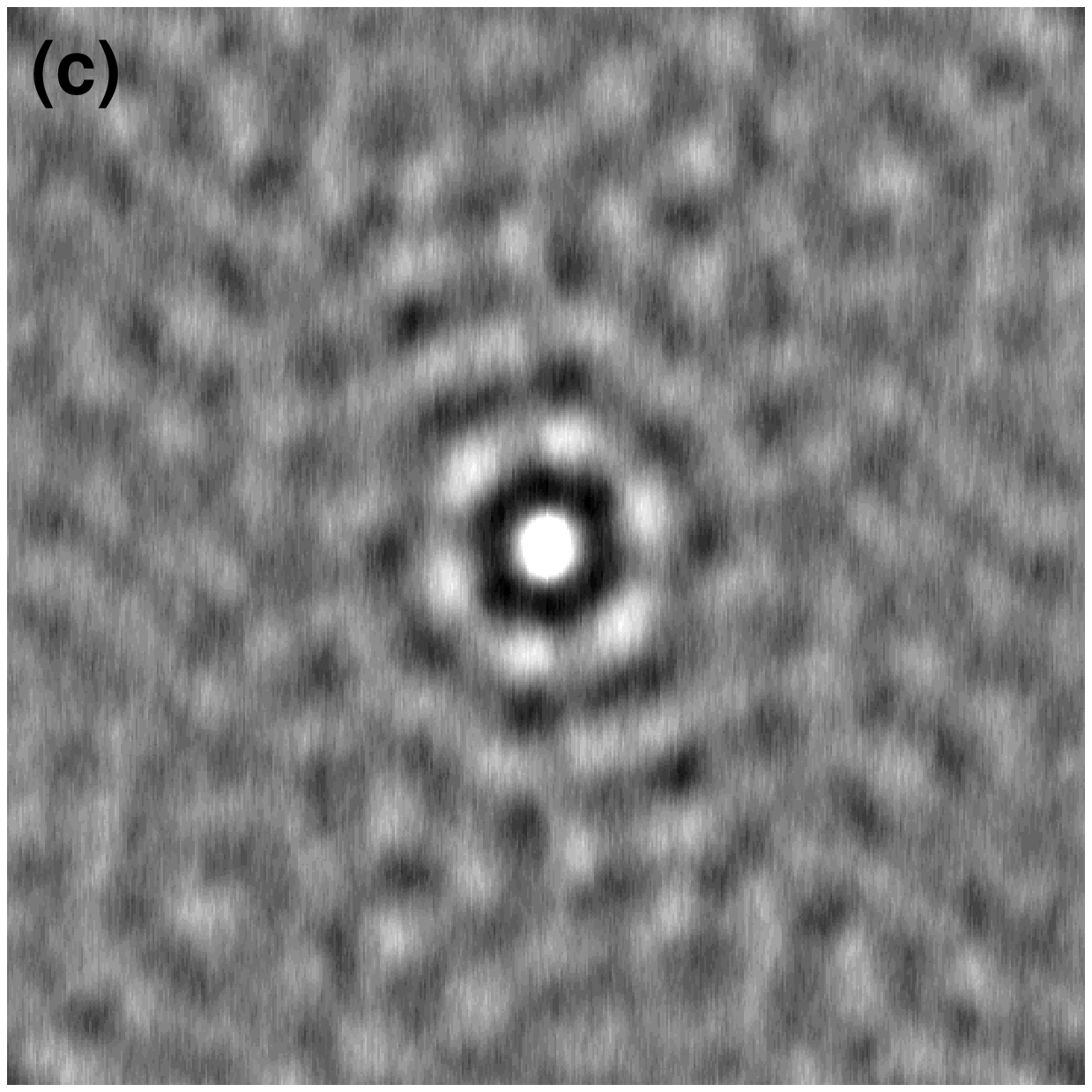}
 \includegraphics[width=0.238\textwidth]{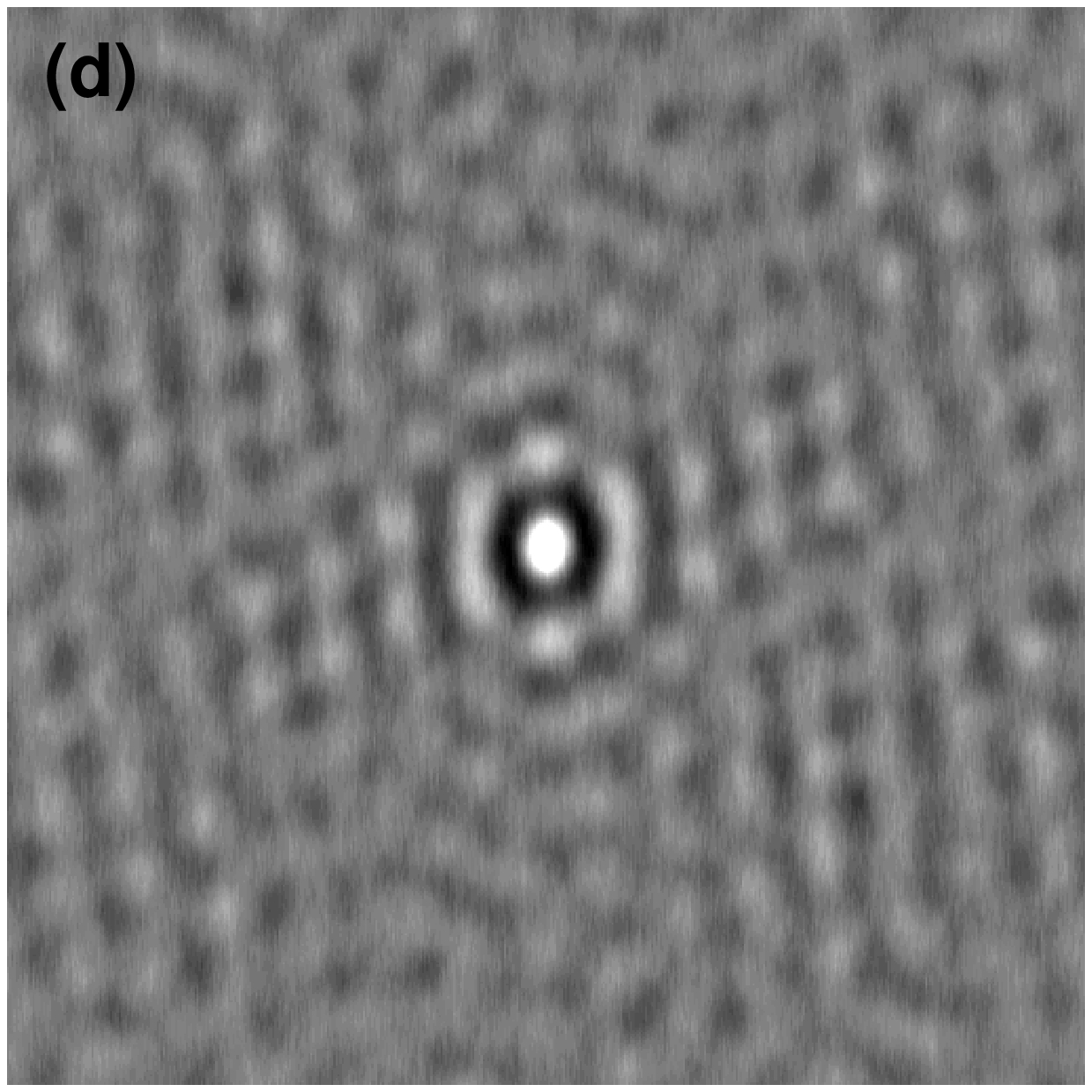}
 \caption{\label{Vortex}Top : vortex images ($1.5 \times 1.5 \; \mu m^2$) at two different magnetic fields : (a) 1200 Oe, (b) 
1900 Oe. Bottom : two dimensional autocorrelation function calculated from the above vortex images.}
\end{figure}

	In a magnetic field, for a type II superconductor a mixed state forms where magnetic vortices each carrying one flux quantum $\phi_0$ penetrate the sample. In the absence of pinning and for an isotropic order parameter, the repulsive interaction between flux lines give rise to a periodic lattice of vortices with a triangular symmetry : the Abrikosov vortex lattice \cite{Abrikosov}. A vortex represents a singularity with the superconducting order parameter suppressed at its center. In the clean limit, i.e. $\xi _0/l \ll 1$ ($\xi _0 = \frac{\hbar v_F}{\pi \Delta (0)}$ is the BCS coherence length) confinement of quasiparticles inside the vortex yields discretized energy levels separated in energy by $\Delta ^2/E_F$ \cite{Caroli}. Such bound states were first observed in $NbSe_2$ as a Zero Bias Conductance (ZBC) peak \cite{Hess}. STM studies of vortices in high temperature superconductors have revealed no ZBC peak but additional quasi-particles states at constant energies above and below the Fermi level \cite{Maggio-Aprile, Pan}. In the less theoretically studied dirty limit case, i.e. $\xi _0/l \gg  1$, a mixing of the core bound states by disorder results in a flat normal-metal-like LDOS predicted \cite{Golubov} and experimentally observed in $Nb_{1-x}Ta_xSe_2$ at $1.3\,K$ \cite{Renner}. The temperature independent vortex size is then equal to the dirty coherence length $\xi_s=(\hbar D / 2 \pi k_B T_c)^{1/2}$, where D is the diffusion coefficient. For our sample, upper critical field measurements yield a Ginzburg-Landau coherence length $\xi_{GL} = 15\; nm$ from the expression $H_{c2} = \phi_0/(2\pi\xi_{GL}^2)$ \cite{Bustarret}. With a mean free path $l\sim 1.3\,nm$ and a BCS coherence length $\xi_0 \sim 240\,nm$ evaluated from the relation $\xi_{GL} = 0.855\sqrt{\xi_0 l}$, boron-doped diamond is clearly in the dirty limit with $\xi_s = (2/\pi) \xi_{GL}\simeq 9.5\,nm$.

	We have imaged the vortex arrangement by combining topography and spectroscopy measurements in a magnetic field applied perpendicular to the diamond film. The sample-tip DC bias was set to a voltage $V$ close to the BCS superconducting coherence peak ($eV = \Delta$). Then, the presence of vortex is revealed by a decrease of $\frac{dI}{dV}$ which gave a dark spot in the differential conductance image. In Fig. \ref{Vortex} (a) and (b) we show two $1.5 \times 1.5\; \mu m^2$ vortex images obtained at about $100 \; mK$ and induced by a field $H$ of 1200 Oe and 1900 Oe respectively. The vortex positions were very stable and insensitive to the scanning. In order to reduce the noise, these images were convoluted with a disc of diameter charateristic of the lentgh scale over which the contrast changes occur in the raw image. Its radius $R$ is not a direct measurement of the coherence length $\xi_s$ because the decrease of the coherence peak begins well before crossing the vortex core \cite{Golubov}. Indeed, we measured $R = 40 \pm  5\, nm$ and $R = 30 \pm  5\, nm$ for a magnetic field of 1200 and 1900 Oe respectively. This dependency can be explained by a field enhanced vortex-vortex interaction \cite{Golubov}.

\begin{figure}
 \includegraphics[width=0.45\textwidth,bb = 35 20 560 400]{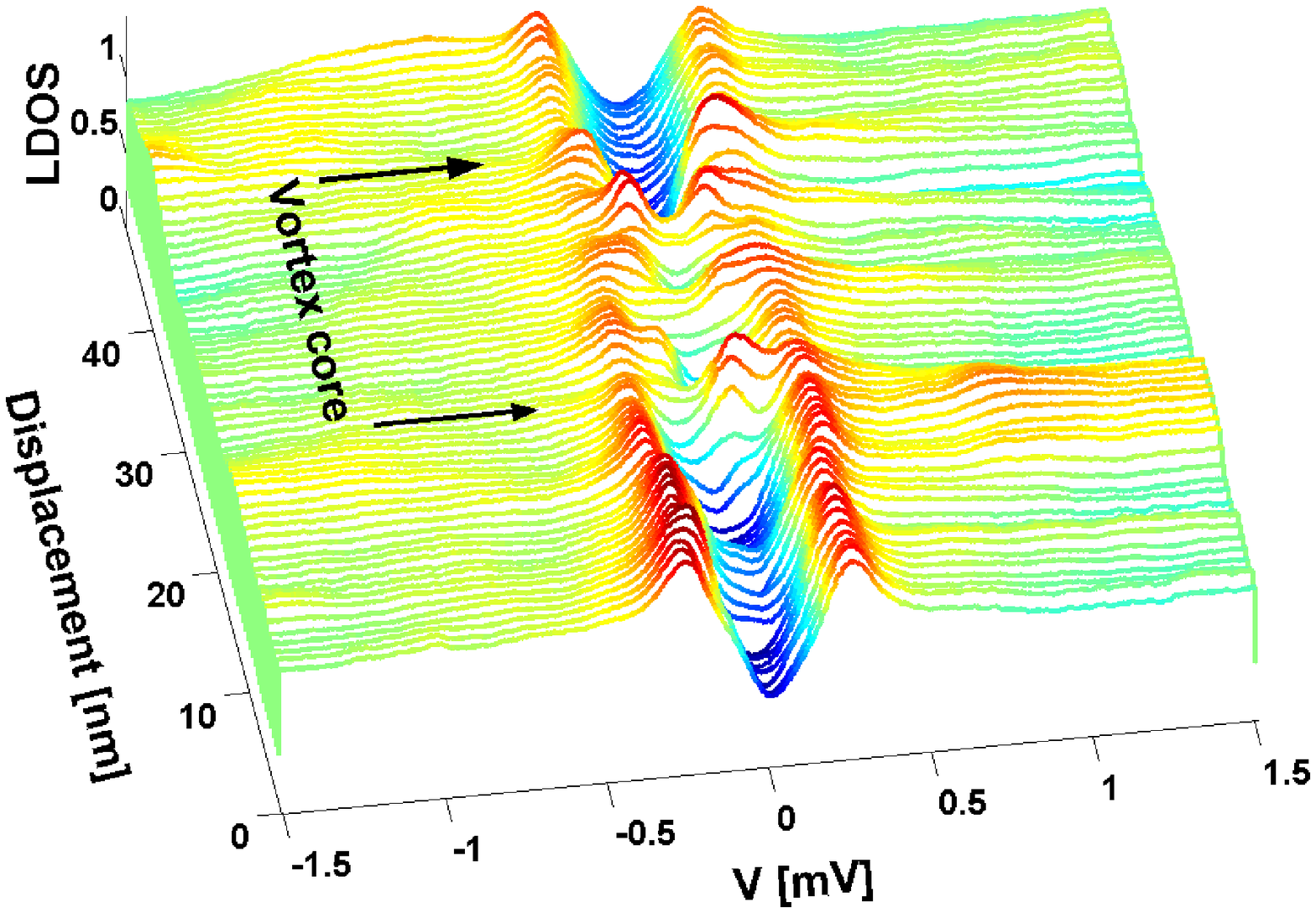}
 \includegraphics[width=0.45\textwidth,bb = 110 260 485 575]{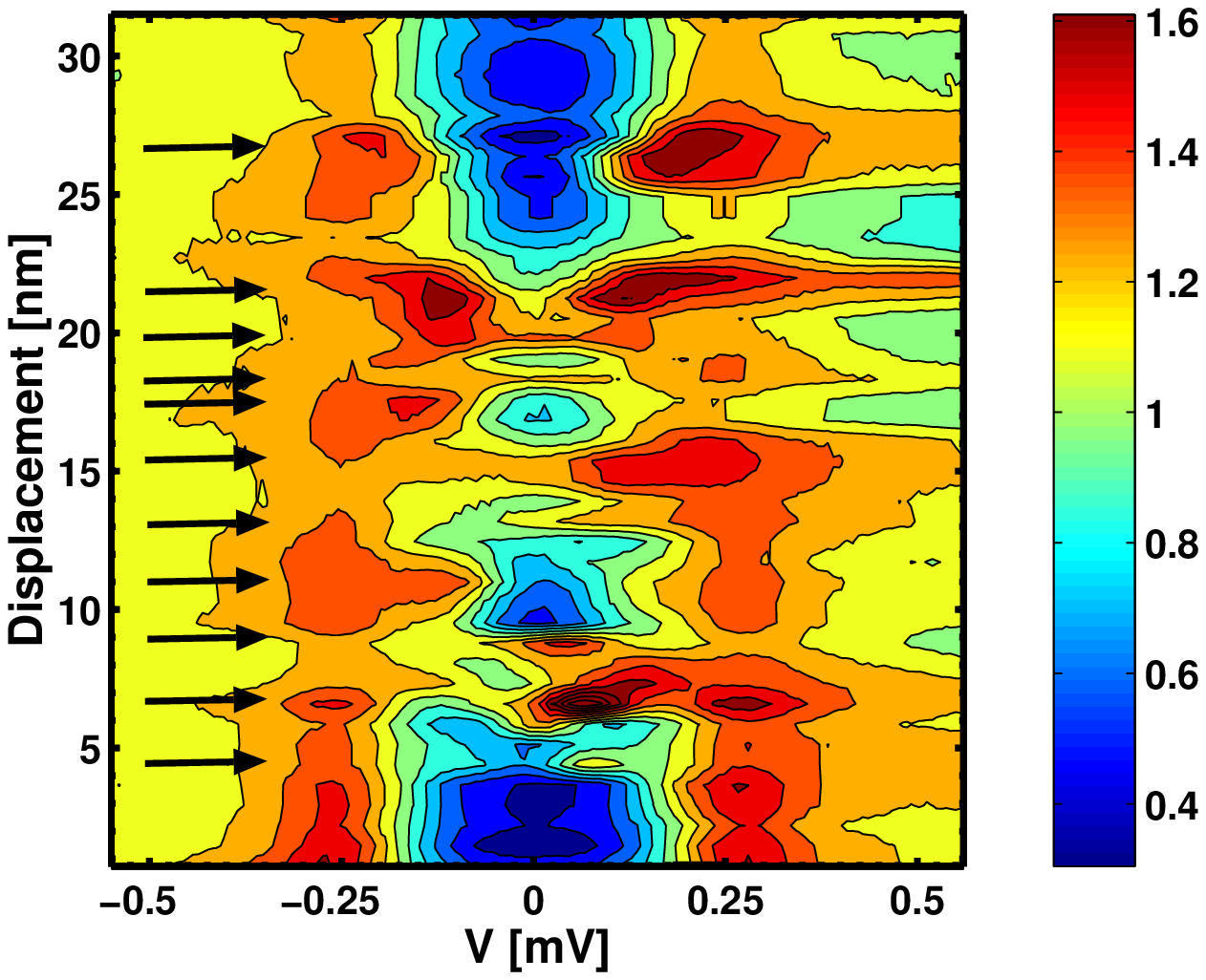} 
 \caption{\label{VortexCore}Tunneling spectroscopy at $T = 53\,mK$ through a vortex core induced by a magnetic field of $1800\,Oe$. Top : tunneling spectra acquired every $0.7\,nm$ along a line crossing the vortex core which extends between the two arrows. A sliding averaging of two consecutive spectra was applied in order to attenuate high peaks for clarity. Bottom : two-dimensional view of the same data restricted to the vortex core and without averaging. The color scale as well as the contour lines give the normalized LDOS. Black arrows indicate the positions of the localized resonances.}
\end{figure}

	 Contrary to the perfect Abrikosov vortex lattice, the spatial distribution of vortices is strongly disordered even though a local hexagonal arrangement persists. In order to characterize it, the autocorrelation function defined by $G(\delta x, \delta y) = \int I(x+\delta x, y+\delta y)I(x,y)\,dx\,dy$ was calculated. Here, $I(x,y)$ is the local degree of brightness. $G(\delta)$ gives the probability of finding a similar brightness in the image for a spatial shift equals to $\delta = (\delta x, \delta y)$ from any point. The results are shown on Fig. \ref{Vortex} (c) and (d) for the vortex images Fig. \ref{Vortex} (a) and (b) respectively. Although the lattice appears strongly disordered, the sixfold symmetry of first rings of neighbors in the autocorrelation pictures indicates a persistent orientational order for any site of the lattice. The mean distance between vortices is equal to $110 \pm 5\; nm$ and $140 \pm 1 \; nm$, in good agreement with $d_\bigtriangleup = (\frac{2}{\sqrt{3}} \frac{\phi _0}{H})^{1/2}$ expected for a triangular lattice. Moreover, for the higher field we can see a rotation and a slight deformation of the ring of first neighbors accompanied by the onset of a diffuse striation in a direction which tends to align with the vicinal stripes of the surface visible on Fig. \ref{Topo}. This alignment is consistent with a stronger pinning by these stripes at higher flux density.

	In order to map the LDOS inside a vortex we acquired spectra every $0.7\,nm$ along lines equally spaced by $2.2\, nm$. The exact localization of the core is a delicate problem in a dirty superconductor since disorder is expected to smear out the ZBC peak by scattering the localized quasiparticule states and consequently yield a flat normal-metal-like LDOS \cite{Renner,Golubov}. Here, instead, we surprisingly obtained numerous localized resonances at non zero energies inside the gap while crossing a vortex along different lines. We consider a vortex core to be defined as the region containing the highest concentration of resonance peaks within our spatial resolution. The corresponding line of spectra is displayed in  Fig. \ref{VortexCore} (top). Such a determination gives a vortex core radius of about $10\,nm$ consistent with $\xi_s$. More striking is the periodic-like spatial occurrence of these resonances as pointed by the black arrows in the two-dimensionnal view of Fig. \ref{VortexCore}. It is worth noticing that a periodic pattern of quasi-particles states has also been observed around vortex cores in $Bi_2Sr_2Ca Cu_2O_{8+\delta}$,      another superconductor close to a Mott transition \cite{Hoffman}. The distance of about $2.2\,nm$ between each localized resonance is almost constant and present on most scanned line. Such a spatial modulation appears in the theoretical clean limit \cite{Hayashi} at low temperature $T/T_c \ll 1$ as Friedel-like oscillations with a period of the order of the Fermi wave-length. Although the distance between each localized resonance is  consistent with an estimation of $\lambda_F \lesssim  1.6\,nm$ in a free-electron model, our sample is not in the clean limit. Moreover, because of thermal smearing, quantum states at a energy of $\Delta ^2/E_F$, would appear as a ZBC peak. On the other hand, these puzzling resonances could be associated to low-lying vortex bound states shifted away from the Fermi level by the impurity potential \cite{Han,Brezin}. Then, the spatial position and the energy of these resonances would be specific to the local configuration of disorder.

	In conclusion, we performed the first spectroscopic tunneling study and STM images of the vortex lattice in superconducting boron-doped diamond. Our results provide a local density of states in excellent agreement with the BCS s-wave theory with a ratio $\Delta/k_B T_c\simeq 1.74$ characteristic of a weak coupling superconductor. Under magnetic field, vortices which are arranged in a strongly disordered triangular lattice present unexpected resonances inside the BCS gap at non zero energies. Further investigations are needed in order to understand whether these magnetic field-induced resonances are a generic feature of any superconducting state neighboring a Anderson-Mott insulator.

\begin{acknowledgments}
	The authors are indebted to C. Cytermann (Technion Haifa) for measuring the SIMS profile of the sample analysed in this work.
\end{acknowledgments}

\end{document}